\documentclass[useAMS,usenatbib,referee]{biom}
\def\bSig\mathbf{\Sigma}

\usepackage[figuresright]{rotating}
\usepackage{nameref}
\usepackage{xcolor}
\usepackage{amsmath}
\usepackage{enumitem}
\usepackage{float}
\usepackage{amsfonts}
\usepackage[authoryear]{natbib}

\title[Accounting for spillover with ASCM]{Accounting for spillover when using the augmented synthetic control method: estimating the effect of localized COVID-19 lockdowns in Chile}

\author{Taylor Krajewski$^{*}$\email{tjk@email.unc.com} \\
	   Department of Biostatistics, University of North Carolina at Chapel Hill, 
Chapel Hill, North Carolina, U.S.A.
	   \and 
	   Michael Hudgens $^{*}$\email{mhudgens@email.unc.edu}\\
	   Department of Biostatistics, University of North Carolina at Chapel Hill, 
Chapel Hill, North Carolina, U.S.A.
	   }

\begin{document}

% \date{{\it Received XXX} 2024. {\it Revised February} 2005.\newline 
% {\it Accepted March} 2005.}

\pagerange{\pageref{firstpage}--\pageref{lastpage}} \pubyear{2024}

\volume{XX}
\artmonth{XXX}
\doi{XXX/XXX.x}

%  This label and the label ``lastpage'' are used by the \pagerange
%  command above to give the page range for the article

\label{firstpage}

%  pub the summary here

\begin{abstract}
The implementation of public health policies, particularly in a single or small set of units (e.g., regions), can create complex dynamics with effects extending beyond the directly treated areas. This paper examines the direct effect of COVID-19 lockdowns in Chile on the comunas where they were enacted and the spillover effect from neighboring comunas. To draw inference about these effects, the Augmented Synthetic Control Method (ASCM) is extended to account for interference between neighboring units by introducing a stratified control framework. Specifically, Ridge ASCM with stratified controls (ASCM-SC) is proposed to partition control units based on treatment exposure. By leveraging control units that are untreated, or treated but outside the treated unit's neighborhood, this method estimates both the direct and spillover effects of intervention on treated and neighboring units. Simulations demonstrate improved bias reduction under various data-generating processes. ASCM-SC is applied to estimate the direct and total (direct + indirect) effects of COVID-19 lockdowns in Chile at the start of the COVID-19 pandemic. This method provides a more flexible approach for estimating the effects of public health interventions in settings with interference.

%\textcolor{red}{Current word count: 182 (needs to stay 225 or less)}

% This approach addresses the challenge of estimating causal effects in settings where neighboring units are also treated, an underexplored issue in synthetic control methodology.

% Our analysis is framed within the potential outcomes framework, and we rely on the Stable Neighbor Treatment Value Assumption (SNTVA) to formally define direct, spillover, and total effects. Simulations illustrate the bias reduction achieved by this method across different scenarios. We apply this approach to estimate the effects of lockdowns on both the treated comunas and their neighbors. This work provides a flexible and robust framework for evaluating public health interventions when interference between units is likely.

% Interference in causal inference arises when the exposure of one unit affects the outcomes of others

% We apply this method to evaluate the impact of localized COVID-19 lockdowns in a small number of comunas in Chile.

\end{abstract}

%
%  Please place your key words in alphabetical order, separated
%  by semicolons, with the first letter of the first word capitalized,
%  and a period at the end of the list.
%

\begin{keywords}
Causal Inference, Interference, Synthetic Controls, COVID-19
\end{keywords}

\maketitle

\section{Introduction}

Synthetic control methods (SCMs) have gained much popularity in panel data settings with a goal of estimating a treatment (or intervention) effect on a single or limited number of treated units. These methods construct a synthetic control, a weighted average of untreated units' outcomes, with weights determined to minimize the difference in pre-treatment outcomes between the treated unit and control units. The synthetic control serves as a proxy for the counterfactual outcome had the treated unit not experienced the intervention, and in turn, the effect of the treatment in the post-treatment period is estimated via the difference between the treated unit's outcome and that of its synthetic control. 

The original SCM (\cite{abadie2003, abadie2010, abadie2021}) was developed for estimating the treatment effect on a single treated unit. Since its inception, SCM has been extended and adapted in a myriad of ways. \cite{abadie2021} provides a detailed overview of many recent SCM-related methodological developments. Many SCM modifications continue to focus on a single treated unit while altering assumptions for greater applicability. For example, regression-based methods that allow for extrapolation (\cite{arkhangelsky2021synthetic, chernozhukov2019, doudchenko2016}), methods that incorporate matrix completion/estimation (\cite{amjad2019, athey2020}), and bias-correction methods (\cite{abadieLH2021, arkhangelsky2021synthetic, ben-michael2021, chernozhukov2019}) have been proposed. Other modifications have extended SCM to account for multiple treated units (\cite{ben-michael2022, abadieLH2021, robbins2017, kreif2016, hainmueller2012}). 

The majority of advancements in synthetic control methodology assume no interference (or spillover) between units, i.e., the treatment of one unit does not affect the outcomes of any other units. However, this assumption may be questionable in settings where units are in close proximity, and units close to the treated unit(s) may experience spillover effects of the treatment. Increasing attention has been paid recently to estimating causal effects in the presence of interference with a goal of disentangling the direct and spillover (or indirect) effects of treatment in both randomized and observational studies \citep{forastiere2021, ogburn2014, tchetgen2012}. However, only a handful of SCM methods allow for spillover effects.

%(Menchetti \& Bojinov 2022; Grossi et al. 2024; Cao \& Dowd 2019; Stefano \& Mellace 2020; Agarwal et al. 2024) 
Some methods have extended SCM to incorporate spillover effects under the partial interference assumption, which limits interference to predefined groups or neighborhoods. For example, \citet{menchetti2022} use a multivariate Bayesian structural time series model to estimate direct and spillover effects, while \citet{grossi2024} apply penalized SCM to estimate synthetic controls for a single treated unit and its neighbors, relying on untreated units outside the neighborhood as controls. Other approaches relax the partial interference assumption but still assume a priori knowledge of units experiencing spillover. \citet{melnychuk2024} developed Iterative SCM to estimate the direct effect on a single treated unit, iteratively removing spillover bias by creating synthetic controls for spillover-affected units. \citet{cao_dowd2019} assume a known linear spillover structure and estimate synthetic controls for all units (treated and controls) but allow all units, including those potentially affected by spillover, to serve as controls in estimating direct and spillover effects. \citet{distefano2024} introduce the inclusive SCM (iSCM) method, which does not require prior knowledge of the spillover structure but does require prior identification of potentially affected units (units that could be indirectly affected by the treatment) and at least one unaffected control for accurate estimation. Their approach allows potentially affected units and other treated units to serve as controls, solving a system of equations to estimate direct and spillover effects.

The work in this manuscript is motivated by a setting in which several units receive treatment, and the units receiving treatment are spatially clustered. While the methods described above focus on estimating the direct effect of treatment on a single treated unit and the spillover effects onto (neighboring) untreated units, our approach estimates the direct effect of treatment on a treated unit as well as the additional spillover effect of a neighboring unit receiving treatment (i.e., spillover between treated units). Specifically, we aim to evaluate the causal effect of localized lockdowns in the initial months of the COVID-19 pandemic on the regions that initially went under lockdown and their neighboring regions that also went under lockdown. 

Methodologically, our approach extends the Augmented Synthetic Control Method (ASCM) \citep{ben-michael2021} by incorporating the Stable Neighbor Treatment Value Assumption (SNTVA), which accounts for interference between neighboring units \citep{agarwal2023}. We develop Ridge ASCM with stratified controls (ASCM-SC) to estimate direct, total, and spillover effects by stratifying control units based on their exposure to treatment. %Similar to \citet{grossi2024}, our method uses a control group of units that are untreated and do not belong to the treated unit’s neighborhood, but these controls are used to estimate the total effect of the intervention, which is the sum of the direct effect and the spillover effect from neighboring treated units. To isolate the direct effect of the intervention on the treated unit of interest, we employ a separate control group of units that are untreated but located in the neighborhood of other treated units. The spillover effect is then estimated as the difference between the total and direct effects of treatment on the unit of interest. 
This approach can improve the estimation of the causal impact of interventions in settings where multiple neighboring units are treated, and spillover effects are likely to occur between treated units.

% Similar to \citet{grossi2024}, the proposed method makes use of a control group of units that are untreated and do not belong to the treated unit’s neighborhood, but these controls are used to estimate the total effect of the intervention (the sum of the direct effect and the spillover effect from neighboring treated units). To estimate the direct effect of the intervention on the treated unit of interest, we employ a separate control group of units that are untreated but located in the neighborhood of a treated unit. The spillover effect is then estimated via the difference between the total and direct effects of treatment on the unit of interest. This approach allows us to more accurately estimate the causal impact of interventions in settings where multiple neighboring units are treated, and spillover effects are likely to occur between treated units.

The paper is organized as follows: Section \ref{sec:chp3_data} provides details of the motivating dataset. Section \ref{sec:paper2_methods} presents the methodology, providing an overview of SCM and ASCM and presenting a modification that accounts for potential spillover effects. In Section \ref{sec:chp3_sims} results from a simulation study are presented, and in Section \ref{sec:chp3_applied_chile} the method is applied to estimate the effects of COVID-19 comuna-level lockdowns in Chile. Section 6 concludes with a discussion.

\section{Motivating Example}\label{sec:chp3_data}

During the first year of the COVID-19 pandemic, the country of Chile implemented certain preventative policies at the national level (e.g., mandatory masks), but localized lockdowns were implemented at the comuna (municipality) level, the smallest regional level in Chile, from March through June 2020. The implementation and length of lockdowns in each municipality and its neighbors varied greatly \citep{li:2022}, providing a unique opportunity to study the effect of lockdowns at the comuna level given the heterogeneous nature of their implementation across comunas and their neighbors. \citet{li:2022} combined administrative COVID-19 surveillance records, a nationally representative household survey, census data, and epidemiologic surveillance records to curate a dataset tracking comuna-level COVID-19 outcomes during this period. COVID-19 cases included symptomatic or asymptomatic SARS-CoV-2 infections confirmed by a positive PCR test. The rich dataset also %collected for this study contains detailed time-series data on COVID-19 case incidence, adjusted for possible reporting lags as described above. Additionally, it 
includes important covariates, such as the proportion of the population in neighboring municipalities under lockdown each day as well as several comuna-level sociodemographic variables that could influence virus transmission: the proportion of the population who were female, living in rural areas, older than 65, living below the official poverty level, living in overcrowded households, and lacking adequate sanitation infrastructure, as well as average monthly income. 

Several adjustments were made to the dataset to ensure the accuracy of the COVID-19 case incidence series \citep{li:2022}. First, incomplete data were imputed by interpolating between the closest available dates with complete data. Given that cumulative case counts were often reported only every two to four days, this imputation step was necessary to fill gaps in the reporting timeline. Second, reporting lags were corrected using the Pruned Exact Linear Time (PELT) algorithm \citep{pelt}. Next, the incidence series was adjusted for space- and time-varying reporting delays, following the methodology of \citet{Zhao_Ju_BETS}, which allowed for reporting lags to vary across comunas and over time, aligning incidence with the day of symptom onset rather than the day of reporting. Finally, the instantaneous reproduction number, the average number of secondary cases per primary infected case, was estimated from the adjusted COVID-19 series using the methodology in \citet{Cori}. %These adjustments were implemented for more accurate estimation of the instantaneous reproduction number ($R_t$) across all municipalities.

%The rich dataset also %collected for this study contains detailed time-series data on COVID-19 case incidence, adjusted for possible reporting lags as described above. Additionally, it 
%includes important covariates, such as the proportion of the population in neighboring municipalities under lockdown each day as well as several comuna-level socio-demographic variables that could influence virus transmission: the proportion of the population who were female, living in rural areas, older than 65, living below the official poverty level, living in overcrowded households, and lacking adequate sanitation infrastructure, as well as average monthly income. %Since only seven comunas initiated lockdown during the first wave of lockdowns in Chile that began at the end of March, 2020, these data provide a valuable foundation for assessing not only the direct effects of lockdowns but also the potential spillover effects when neighboring municipalities are under lockdown.  %This motivates the use of advanced methodological approaches that account for potential spillover effects when estimating counterfactuals.

Since only seven comunas initiated lockdown during the first wave of lockdowns in Chile that began at the end of March 2020, these data provide a valuable foundation for assessing not only the direct effects of lockdowns but also the potential spillover effects when neighboring municipalities are under lockdown. \citet{li:2022} used this data to estimate the effect of extending lockdown duration in specific comunas and their neighbors. Their study focused on the first wave of lockdowns in Chile that began at the end of March 2020, aiming to quantify the impact of keeping a comuna (and additionally its neighbors) under lockdown versus lifting the restrictions. Further details on this study are discussed in Section \ref{sec:chp3_Yige} and Web Appendix A. In contrast, the analysis in this paper (presented in Section \ref{sec:chp3_applied_chile}) treats the initial lockdown implementation as the primary intervention of interest. ASCM with stratified controls (ASCM-SC), as described in Section \ref{sec:chp3_ascm-sc}, is implemented to estimate the direct effect of a comuna going under lockdown on the instantaneous rate of transmission of COVID-19. Additionally, ASCM-SC estimates the total effect when accounting for the effect of some proportion of neighboring comunas going under lockdown, and in consequence, the indirect effect of neighboring comunas undergoing lockdown. 

\section{Methodology}\label{sec:paper2_methods}

\subsection{Overview of SCM and ASCM}\label{overview_scm_ascm}
\subsubsection{SCM}
Begin with a typical SCM setup, where units $i=1,\ldots,N$ are observed for $t=1,\ldots,T$ time periods and $N$ is fixed. Suppose only unit $i=1$ receives treatment (the setting with multiple treated units is considered below). Let units $i=2,\ldots,N$ serve as the control units which are untreated throughout the $T$ time periods, often referred to as the ``donor pool." The treated unit is unexposed at time points $t=1,\ldots, T_0$ (the pre-treatment period) and exposed for $t>T_0$.

 Let $A_{it}$ be a treatment indicator for unit $i$ at time $t$; i.e., a dichotomous random variable having values 0 or 1. Assuming no interference between units, the potential outcome for unit $i$ at time $t$ is denoted by $Y_{it}(a)$ for $a=\{0,1\}$. Let
 % Let $a_{it}$ denote possible values of $A_{it}$. The potential outcomes for unit $i$ at time $t$ are written as $Y_{it}(a_{it}=a)$, which for simplicity can be written as $Y_{it}(a)$. Letting $W_{it}$ be an indicator of treatment for unit $i$ at time $t$, 
 the observed outcome for unit $i$ at time $t$ be $Y_{it}=A_{it}Y_{it}(1)+(1-A_{it})Y_{it}(0)$. The goal is to estimate the effect of the intervention on the treated unit: $\tau_{1t}=Y_{1t}(1)-Y_{1t}(0)=Y_{1t}-Y_{1t}(0)$, for $t>T_0$. %, where the second equality holds by consistency. 
 Since $Y_{1t}(1)$ is observed for $t>T_0$, $Y_{1t}(0)$ must be estimated in order to estimate $\tau_{1t}$.

The inaugural SCM (\cite{abadie2015, abadie2010, abadie2003}) lets $\boldsymbol{\gamma}=(\gamma_2,\ldots,\gamma_N)'$ be a vector of weights subject to the constraints
\begin{equation}\label{eq:2SCM_constraints_greater_0}
    \gamma_i \geq 0 \;\text{for}\; i=2,\ldots,N, \text{ and } 
\end{equation} 
\begin{equation}\label{eq:2SCM_constraints_sum}
     \sum \gamma_i=1,
\end{equation}
\noindent which limit $\boldsymbol{\gamma}$  to the simplex $\triangle^{N_0} = \{\boldsymbol{\gamma} \in \mathbb{R}^{N_0} | \gamma_i \geq 0 \: \forall i, \sum\gamma_i =1\}$. Throughout, $\sum$ indicates $\sum_{i=2}^N$. Each weight vector $\boldsymbol{\gamma}$ can be used to construct an estimator of $Y_{1t}(0)$ given by $\hat{Y}_{1t}(0)=\sum\gamma_i Y_{it}$. The SCM estimator corresponds to weights $\hat{\boldsymbol{\gamma}}^{\text{scm}}$ which minimize differences in pre-treatment period data between the treated unit and control units by solving the optimization problem, 
\begin{equation}\label{eq:2SCM_weight_minimizer}
    \underset{\substack{\gamma}}{\mbox{min}}||\boldsymbol{V}^\frac{1}{2}(\mathbf{X}_1-\mathbf{X}_0'\boldsymbol{\gamma})||_2^2= \underset{\substack{\gamma}}{\mbox{min}}[(\mathbf{X}_1-\mathbf{X}_0'\boldsymbol{\gamma})'\boldsymbol{V}(\mathbf{X}_1-\mathbf{X}_0'\boldsymbol{\gamma})],
\end{equation}

\noindent subject to constraints (\ref{eq:2SCM_constraints_greater_0}) and (\ref{eq:2SCM_constraints_sum}). In (\ref{eq:2SCM_weight_minimizer}) the matrix $\boldsymbol{V}$ is specified by the analyst, and the vector $\mathbf{X}_1$ and matrix $\mathbf{X}_0$ contain pre-treatment covariates and outcomes for the treated unit and control units, respectively. In particular, $\mathbf{X}_1= \nolinebreak(\mathbf{Z}'_1, \mathbf{Y}'_1)'$ is the vector describing the treated unit ($i=1$) in the pre-treatment period, and  $\mathbf{X}_0$ is the matrix describing the control units ($i=2,\ldots,N$) in the pre-treatment period such that the $i$th column of $\mathbf{X}_0$ is $\mathbf{X}_i = (\mathbf{Z}'_i, \mathbf{Y}'_i)'$, where $\mathbf{Z}_i$ is a (column) vector of covariates %(not affected by the treatment) assumed to be predictive of the outcome variable 
measured during the pre-treatment period, and $\mathbf{Y}_i$ is the vector of observed outcomes during the pre-treatment period.  The importance matrix, $\boldsymbol{V}$, is a symmetric, positive semidefinite matrix that can be chosen a priori (e.g., $\boldsymbol{V}$ may be the identity matrix) or based on the observed data \citep{abadie2021, abadie2015, abadie2010}. The SCM estimate of $\tau_{1t}$ is thus $\hat{\tau}_{1t}^{\text{scm}}=Y_{1t}-\sum\hat{\gamma}^{\text{scm}}_iY_{it}$.

\subsubsection{ASCM}
SCM can be used to estimate the effect of an intervention on a single treated unit. The constraints (\ref{eq:2SCM_constraints_greater_0}) and (\ref{eq:2SCM_constraints_sum}) allow for simple interpretation of the counterfactual (i.e., typically a few control units contribute to the counterfactual with varying, positive weights). However, SCM is not recommended in settings with a small number of pre-treatment periods or in the case of poor pre-treatment fit (\cite{abadie2021, abadie2015}). \citet{abadie2010} demonstrated that in cases where the treated unit's vector of pre-treatment covariates and lagged outcomes, $\mathbf{X}_1$, can be matched exactly by a weighted combination of the control units' covariates and outcomes, $\mathbf{X}_0\hat{\bm{\gamma}}^{\text{scm}}$, (i.e., when $\mathbf{X}_1 = \mathbf{X}_0\hat{\bm{\gamma}}^{scm}$), the SCM estimator is unbiased under an autoregressive model and is asymptotically unbiased as the number of pre-treatment periods increases under a linear factor model. However, as highlighted by \citet{ben-michael2021}, often it is not possible to find $\hat{\bm{\gamma}}^{\text{scm}}$ such that $\mathbf{X}_1 = \mathbf{X}_0\hat{\bm{\gamma}}^{\text{scm}}$; in other words, it can be challenging to obtain perfect pre-treatment fit. %the curse of dimensionality often prevents the treated unit from lying inside the convex hull of control units' pre-treatment values, %(i.e., $\mathbf{X}_1 = \mathbf{X}_0\bm{\gamma}^{\text{scm}}$), making it challenging to achieve perfect pre-treatment fit. 
In such scenarios, SCM is not recommended due to potential bias. %Thus, strategies that improve pre-treatment fit can reduce the bias of the SCM estimator. 

% However, SCM is not recommended in settings with a small number of pre-treatment periods or in the case of poor pre-treatment fit (\cite{abadie2015, abadie2021}). Under various data-generating processes (DGPs), the bias of the SCM estimator relies on $\mathbf{X}_1 = \mathbf{X}_0^{scm}$ (i.e., the synthetic control perfectly fits the treated unit in the pre-treatment period), and thus improving pre-treatment fit reduces the bias of the SCM estimator. 

The Augmented SCM (ASCM) modifies the SCM approach to adjust for poor pre-treatment fit to ultimately reduce the bias of the SCM estimator (\cite{krajewski2024, ben-michael2021}). ASCM corrects the SCM estimator using a model-based estimate of bias. Suppose $Y_{it}(0)=m_{it}+\epsilon_{it}$, where $m_{it}$ may be a function of pre-treatment outcomes, $\mathbf{Y}_i$, and/or covariates predictive of the outcome, $\mathbf{Z}_i$, and the error terms $\epsilon_{it}$ are sub-Gaussian assumed to have mean zero. %with scale parameter $\sigma$. 
Letting $\hat{m}_{it}$ be an estimator of $m_{it}$ for $t>T_0$, %and limiting to one post-treatment period $\boldsymbol{T}$ for simplicity, 
the  ASCM estimator of $Y_{1t}(0)$ for $t>T_0$ is
\begin{equation} \label{eq:ASCM_estimator}
    \hat{Y}_{1t}^{\text{aug}}(0)=\sum\hat{\gamma}_i^{\text{scm}}Y_{it}+\hat{m}_{1t}-\sum\hat{\gamma}_i^{\text{scm}}\hat{m}_{it}.
\end{equation}

The ASCM estimator is thus the original SCM estimator with an added term, $\hat{m}_{1t}-\nolinebreak\sum\hat{\gamma}_i^{\text{scm}}\hat{m}_{it}$, that corrects for the model-based estimate of bias. While various estimators of $\hat{m}_{it}$ could be used, ridge regression is recommended (\cite{ben-michael2021}). Substituting  $\hat{m}_{it}=\hat{\eta}_0^{\text{ridge}}+\mathbf{X}_i'\hat{\boldsymbol{\eta}}^{\text{ridge}}$, where $\hat{\eta}_0^{\text{ridge}}$ and $\hat{\boldsymbol{\eta}}^{\text{ridge}}$ solve $\underset{\substack{\eta_0, \boldsymbol{\eta}}}{\mbox{arg min}}\frac{1}{2}\sum(Y_i-(\eta_0+\mathbf{X}_i'\boldsymbol{\eta}))^2+\lambda^{\text{ridge}}||\boldsymbol{\eta}||_2^2$ into equation (\ref{eq:ASCM_estimator}) results in the Ridge ASCM estimator $\hat{Y}_{1t}^{\text{aug}}(0)=\sum\hat{\gamma}_i^{\text{scm}}Y_{it}+\bigg(\mathbf{X}_1-\sum\hat{\gamma}_i^{\text{scm}}\mathbf{X}_i\bigg)\hat{\boldsymbol{\eta}}^{\text{ridge}}$, for $t>T_0$, or equivalently, $\hat{Y}_{1t}^{\text{aug}}(0)=\sum{\hat{\gamma}_i}^{\text{aug}}Y_{it}$,
%\begin{equation}\label{eq:Ridge_ASCM_est}
%\end{equation}
%\noindent or equivalently,
% \begin{equation}\label{eq:ridge_ASCM_est2}
% \end{equation}
where ${\hat{\gamma}_i}^{\text{aug}}=\hat{\gamma}_i^{\text{scm}}+(\mathbf{X}_1-\mathbf{X}_0'\hat{\boldsymbol{\gamma}}^{\text{scm}})'(\mathbf{X}_0'\mathbf{X}_0+\lambda^{\text{ridge}}\bm{I}_{T_0})^{-1}\mathbf{X}_i$,
% \begin{equation}\label{eq:weights_aug}
% \end{equation}
% \noindent 
with $\bm{I}_{T_0}$ the $T_0 \times T_0$ identity matrix. In this form, the Ridge ASCM weights ${\hat{\boldsymbol{\gamma}}}^{\text{aug}}$ are the solution to
\begin{equation}\label{eq:Ridge_ASCM_weight_minimizer}
    \underset{\substack{\gamma}}{\mbox{min}}\frac{1}{2\lambda^{\text{ridge}}}||\mathbf{X}_1-\mathbf{X}_0'\boldsymbol{\gamma}||_2^2+\frac{1}{2}||\boldsymbol{\gamma}-\hat{\boldsymbol{\gamma}}^{\text{scm}}||_2^2,
\end{equation}

\noindent subject to constraint (\ref{eq:2SCM_constraints_sum}) on $\boldsymbol{\gamma}$.

% \begin{equation}\label{eq:2Y1t_estimator}
%     \hat{Y}_{1t}(0)=\sum\gamma_i Y_{it}.
% \end{equation}
% \begin{equation}\label{eq:2tau_hat}
%     \hat{\tau}_{1t}^{\text{scm}}=Y_{1t}-\sum\gamma^{\text{scm}}_iY_{it}.
% \end{equation} 

\subsection{Accounting for spillover}\label{sec:chp3_ascm-sc}

The above, standard (A)SCM setup presumes the Stable Unit Treatment Value Assumption (SUTVA), which supposes each unit's outcome is not affected by the treatment of other units and there are not multiple versions of treatment (\cite{rubin1980}). As a consequence, all untreated units comprise the donor pool. However, when treatment is community lockdown during a pandemic and the unit of interest is a geographic region (for example, a comuna), multiple units may receive treatment while many control units could be indirectly affected by neighboring regions' interventions. If one were to apply ASCM in this setting, the estimated treatment effects could be biased. This bias might occur in two ways. First, ASCM might attribute the impact of neighboring regions' interventions to the treated unit's outcome. Second, it might count the indirect effects on control units as part of the treatment effect. Specifically, if control units are indirectly affected by neighboring lockdowns, the difference in outcomes between treated and control regions reflects both the direct impact and these spillover effects. As a result, the estimated treatment effect becomes a blend of the true treatment impact and the indirect influences, potentially leading to an over- or underestimation of the actual effect on the treated unit.

%, either by incorporating spillover effects from other treated units and/or by conflating the spillover effects experienced by control units with the estimated treatment effect on the treated unit.

%Additionally, potential control units under the standard SCM setup include only those units that do not experience the same or similar versions of treatment that the unit of interest experiences. 

In this or similar scenarios, it is important to consider not only the impact of treatment on the treated unit of interest but also the impact of neighboring units' treatment on the unit of interest. To do so, the focus shifts beyond estimating the direct effect, $\tau_{it}=Y_{it}(1)-Y_{it}(0)=Y_{it}-Y_{it}(0)$, under the assumption of no interference, to instead estimating the direct, total, and in consequence, spillover effects of treatment on the unit of interest. This section outlines the implementation of Ridge ASCM with stratified controls (ASCM-SC) in which control units are stratified to separately estimate the direct and total effects of the treatment on the treated unit.

\subsubsection{Notation}\label{sec:paper2_notation}

Continue with the notation from Section \ref{overview_scm_ascm} where $N$ units, indexed by $i=1,\ldots,N$ are each observed for $T$ time periods $t=1,\ldots,T$. Assume $N$ is fixed but do not assume there is a single treated unit. %Assume there are not multiple versions of treatment (\cite{rubin1980}), and no anticipation of treatment (i.e., the intervention has no effect on the outcome prior to implementation) (\cite{abadie2010}).
Let $\boldsymbol{a}=(a_1, \ldots, a_N)$ be the vector of treatments across all $N$ units, where $a_i$ denotes the treatment received by unit $i$. 
Additionally, let $\mathcal{N}(i)$ denote the neighbors of unit $i$ (spatially, this includes units whose borders touch unit $i$; in terms of a network, this includes units directly connected to unit $i$). Assume the potential outcome of unit $i$ at time $t$, denoted $Y_{it}(\boldsymbol{a})$, may depend on $i$'s own treatment $a_i$ and the treatments of its neighbors' $\boldsymbol{a}_{\mathcal{N}(i)}$, but not on the treatment of any other unit. \cite{agarwal2023} refer to this as the Stable Neighbor Treatment Value Assumption (SNTVA). Formally, under SNTVA, $Y_{it}(\boldsymbol{a}) = Y_{it}(a_i, \boldsymbol{a}_{\mathcal{N}(i)})$.

%the potental outcome for unit $i$ at time $t$ under treatment $a$ is given by $Y_{it}(a) = Y_{it}(a, \boldsymbol{a}_{\mathcal{N}(i)})$, where $\boldsymbol{a}_{\mathcal{N}(i)}$ denotes the treatment of units in unit $i$'s neighborhood $\mathcal{N}(i)$ at time $t$. 

% \noindent where $\boldsymbol{a}_{\mathcal{N}(i)}$ denotes the treatment of units in unit $i$'s neighborhood $\mathcal{N}(i)$ at time $t$

% \bigskip

% \noindent \textbf{Assumption 1} (Stable Neighbor Treatment Value Assumption [SNTVA]). The potental outcome for unit $i$ at time $t$ under treatment $a$ is given by
% \begin{equation} \nonumber
%     Y_{it}(a) = Y_{it}(a, \boldsymbol{a}_{\mathcal{N}(i)}),
% \end{equation}

% \noindent where $\boldsymbol{a}_{\mathcal{N}(i)}$ denotes the treatment of units in unit $i$'s neighborhood $\mathcal{N}(i)$ at time $t$. 

% \bigskip

Let $q_{(i)t}=f(\boldsymbol{a}_{\mathcal{N}(i)})$ be a binary function that summarizes the treatment vector of $\mathcal{N}(i)$ into a single variable. For example, $q_{(i)t}$ could be an indicator of whether or not any proportion of $\mathcal{N}(i)$ are treated, i.e. $q_{(i)t}= 0$ if $\boldsymbol{a}_{\mathcal{N}(i)} = [0,0,\ldots,0]' \text{ at time }t$ and $q_{(i)t}= 1$ otherwise.
% \[ q_{(i)t} = \begin{cases} \nonumber
%       0 & \text{if }\boldsymbol{a}_{\mathcal{N}(i)} = [0,0,\ldots,0]' \text{ at time }t\\
%       1 & \text{otherwise} \\
%    \end{cases}
% \]
Now, further assume that the potential outcome of unit $i$ with neighbors $\mathcal{N}(i)$ is given by $Y_{it}(a_{it}=a, q_{(i)t}=q)$, which for simplicity can be written as $Y_{it}(a, q)$. 

For each treated unit for $t>T_0$, we aim to estimate three effects: the  \emph{Direct Effect}, \emph{Total Effect}, and \emph{Spillover Effect}. The \emph{Direct Effect}, $\tau_{it}^{DE}= Y_{it} (1, 1)-Y_{it}(0, 1) = Y_{it}-Y_{it}(0, 1)$  is the effect of unit $i$ receiving treatment when at least one neighbor is also treated. The \emph{Total Effect} is the effect of unit $i$ receiving the intervention in addition to some proportion of $\mathcal{N}(i)$ receiving the intervention, i.e.$\tau_{it}^{TE}= Y_{it}(1, 1)-Y_{it}(0, 0) = Y_{it}-Y_{it}(0, 0)$. The \emph{Spillover Effect} is the portion of the total effect attributable solely to neighbors being treated, or the effect on unit $i$ of some proportion of $\mathcal{N}(i)$ receiving the intervention, computed as $\tau_{it}^{SE}= \tau_{it}^{TE}-\tau_{it}^{DE} = \bigg(Y_{it}(1, 1)-Y_{it}(0, 0)\bigg) - \bigg(Y_{it}(1, 1)-Y_{it}(0, 1)\bigg) = Y_{it}(0, 1)-Y_{it}(0, 0)$.

Note that the second equality in the definitions of the direct and total effects holds by consistency, since $Y_{it}(1,1)$ is observed, if at least one of unit $i$'s neighbors is treated. Additionally, a second type of direct effect could be defined, $ Y_{it} (1, 0)-Y_{it}(0, 0)$, to measure the effect of unit $i$ receiving treatment when no neighbors are treated. However, this direct effect is only applicable in scenarios where units exist that are treated with no treated neighbors (i.e., when $Y_{it}(1,0)$ can be estimated).

    % \begin{enumerate}[leftmargin=1cm]
    %     \item \textbf{Direct Effect} (the effect of the intervention on unit $i$): 
    %     \begin{equation}
    %     \label{eqn:2directeffect}
    %     \tau_{it}^{DE}= Y_{it}(1, 1)-Y_{it}(0, 1) = Y_{it}-Y_{it}(0, 1),
    %     \end{equation}
        
    %     %\noindent where the second equality holds by consistency since $Y_{it}(1,1)$ is observed.
        
    %     \bigskip

    %     \item \textbf{Total Effect} (the effect of unit $i$ receiving the intervention in addition to some proportion of $\mathcal{N}(i)$ receiving the intervention):
    %     \begin{equation}
    %     \label{eqn:2totaleffect}
    %     \tau_{it}^{TE}= Y_{it}(1, 1)-Y_{it}(0, 0) = Y_{it}-Y_{it}(0, 0),
    %     \end{equation}

    %     \noindent where the second equality in (\ref{eqn:2directeffect}) and (\ref{eqn:2totaleffect}) holds by consistency since $Y_{it}(1,1)$ is observed, and
        
    %     \bigskip
 
    %     \item \textbf{Spillover Effect} (the effect on unit $i$ of some proportion of $\mathcal{N}(i)$ receiving the intervention):
    %     \begin{equation}
    %     \label{eqn:2spillovereffect}
    %     \tau_{it}^{SE}= \tau_{it}^{TE}-\tau_{it}^{DE} = \bigg(Y_{it}(1, 1)-Y_{it}(0, 0)\bigg) - \bigg(Y_{it}(1, 1)-Y_{it}(0, 1)\bigg) = Y_{it}(0, 1)-Y_{it}(0, 0)
    %     \end{equation}

    % \end{enumerate}

\subsubsection{Control Units}

In a setting where several units simultaneously receive an intervention,  units fall into one of four categories at time $t>T_0$. First, let $\mathcal{S}_{11}$ be the set of treated units with at least one treated neighbor (i.e., $Y_{it}=Y_{it}(1,1)$). Second, let $\mathcal{S}_{10}$ be the set of treated units with no treated neighbors (i.e., $Y_{it}=Y_{it}(1,0)$). Third, let $\mathcal{S}_{01}$ be the set of untreated units with at least one treated neighbor (i.e., $Y_{it}=Y_{it}(0,1)$). Fourth, let $\mathcal{S}_{00}$ be the set of untreated units with no treated neighbors (i.e., $Y_{it}=Y_{it}(0,0)$), sometimes referred to as pure units/controls.  

% \begin{enumerate}[label=(\Alph*),leftmargin=1cm]
%     \item treated units with at least one treated neighbor  [i.e., $Y_{it}=Y_{it}(1,1)$]
%     \item treated units with no treated neighbors [i.e., $Y_{it}=Y_{it}(1,0)$]
%     \item untreated units with at least one treated neighbor [i.e., $Y_{it}=Y_{it}(0,1)$]
%     \item untreated units with no treated neighbors [i.e., $Y_{it}=Y_{it}(0,0)$] \newline (sometimes referred to as pure units/controls) 
% \end{enumerate}

% \bigskip

% Without loss of generality, let $i=1,\ldots,N_A$ be the units in category (A); let $i=N_A+1,\ldots,N_B$ be the units in category (B); let $i=N_B+1,\ldots,N_C$ be the units in category (C); and let $i=N_C+1,\ldots,N$ be the units in category (D). In order to estimate the direct and total effects of the intervention on any unit $i$ in category (A) under Assumption 1 (SNTVA), units not in $\mathcal{N}(i)$ that fall into categories (C) and (D), respectively, are utilized as the donor pool. For example, to estimate the direct effect on unit $i=1$, the donor pool is comprised of $i \in \{N_B+1,\ldots, N_C\}$ such that $i \not \in \mathcal{N}(1)$, and thus are not subject to spillover effects from unit $1$ under Assumption 1. To estimate the total effect on unit $i=1$, the donor pool is comprised of $i \in \{N_C+1,\ldots, N\}$; since $q_{(i)t}$ is an indicator of whether some proportion of unit $i$'s neighbors is exposed, then all units in $\mathcal{N}(1)$ will have $q_{(i)t}=1$ and thus do not fall into category (D). 

\subsubsection{Estimators of Direct, Total, and Spillover Effects}

In order to estimate the direct and total effects of the intervention on any unit $i \in \mathcal{S}_{11}$ under the assumption that the potential outcome of unit $i$ with neighbors $\mathcal{N}(i)$ is given by $Y_{it}(a, q)$, ASCM is implemented utilizing units not in $\mathcal{N}(i)$ that fall into sets $\mathcal{S}_{01}$ and $\mathcal{S}_{00}$, respectively, as the donor pools.

To estimate the direct effect for each unit $i\in\mathcal{S}_{11}$, the donor pool is comprised of all units in $\mathcal{S}_{01}$ that are not in $\mathcal{N}(i)$, and thus are not subject to spillover effects from unit $i$; i.e.,
\begin{equation*}
     \hat{Y}_{it}^{\text{aug}}(0,1)=
\sum\limits_{j \in \{\mathcal{S}_{01}\setminus \mathcal{N}(i)\}}{\gamma_{j}}^{(i)_{DE}} Y_{jt},
\end{equation*}

\noindent where $\gamma_{j}^{(i)_{DE}}$ are weights found via \eqref{eq:Ridge_ASCM_weight_minimizer} with $\mathbf{X}_0$  a matrix describing the units $j \in \{ \mathcal{S}_{01}\setminus \mathcal{N}(i)\}$ such that, for each $i \in \mathcal{S}_{11}$, $\sum\limits_{j \in \{\mathcal{S}_{01} \setminus \mathcal{N}(i)\}
}{\gamma_{j}}^{(i)_{DE}}=1$. For each unit $i \in \mathcal{S}_{11}$, the set of weights $\boldsymbol{\gamma}^{(i)_{DE}}$ %=\Bigg[\gamma_{N_c+1}^{(i)_{DE}},\ldots, \gamma_{N}^{(i)_{DE}}\Bigg]$ 
applied to the units $j \in \{\mathcal{S}_{01} \setminus \mathcal{N}(i)\}$ defines the synthetic control for unit $i$ that serves as a proxy for the counterfactual outcome had unit $i$ not experienced the intervention, but some proportion of units in $\mathcal{N}(i)$ did. In turn, the direct effect is estimated for $t>T_0$ via 
\begin{equation}\label{eqn:2DE}
        \hat{\tau}_{it}^{DE}= Y_{it} - \hat{Y}_{it}^{\text{aug}}(0,1).
\end{equation}

Similarly, to estimate the total effect for each unit $i\in\mathcal{S}_{11}$, the donor pool is comprised of all units in $\mathcal{S}_{00}$; i.e.,
\begin{equation*}
     \hat{Y}_{it}^{\text{aug}}(0,0)=
\sum\limits_{j \in \mathcal{S}_{00}}{\gamma_{j}}^{(i)_{TE}} Y_{jt},
\end{equation*}
\noindent where $\gamma_{j}^{(i)_{TE}}$ are weights found via \eqref{eq:Ridge_ASCM_weight_minimizer} with $\mathbf{X}_0$  a matrix describing the units $j \in  \mathcal{S}_{00}$ such that, for each $i \in \mathcal{S}_{11}$, $\sum\limits_{j \in \mathcal{S}_{00}
}{\gamma_{j}}^{(i)_{TE}}=1$. For each unit $i \in \mathcal{S}_{11}$, the set of weights $\boldsymbol{\gamma}^{(i)_{TE}}$ %=\Bigg[\gamma_{N_c+1}^{(i)_{DE}},\ldots, \gamma_{N}^{(i)_{DE}}\Bigg]$ 
applied to the units $j \in \mathcal{S}_{00}$ defines the synthetic control for unit $i$ that serves as a proxy for the counterfactual had the intervention not been experienced by unit $i$ nor units in $\mathcal{N}(i)$. In turn, the total effect is estimated for $t>T_0$ via 
    \begin{equation}
        \label{eqn:2TE}
        \hat{\tau}_{it}^{TE}= Y_{it} - \hat{Y}_{it}^{\text{aug}}(0,0).
    \end{equation}

% To construct synthetic controls that most closely match the treated units in the pre-treatment period, ASCM is employed twice: once to estimate $\boldsymbol{\gamma}^{(i)_{DE}}$ and thus the direct effect, and once to estimate $\boldsymbol{\gamma}^{(i)_{TE}}$ and thus the total effect. For each unit $i \in \{1,\ldots,N_A\}$, given ASCM estimates of weights $\boldsymbol{\hat{\gamma}}^{(i)_{DE}}$ and $\boldsymbol{\hat{\gamma}}^{(i)_{TE}}$, the direct and total effects are estimated for $t>T_0$ as follows:
%     \begin{equation}
%         \label{eqn:2DE}
%         \hat{\tau}_{it}^{DE}= Y_{it} - \sum\limits_{j \in (\{N_B+1, \ldots, N_C\} \setminus \mathcal{N}(i))}{\gamma_{j}}^{(i)_{DE}} Y_{jt}
%     \end{equation}

% \noindent and
%     \begin{equation}
%         \label{eqn:2TE}
%         \hat{\tau}_{it}^{TE}= Y_{it} - \sum\limits_{j\in \{N_C+1,\ldots,N\}}{\gamma_{j}}^{(i)_{TE}} Y_{jt}.
%     \end{equation}

The above estimates (\ref{eqn:2DE}) and (\ref{eqn:2TE}) can then be subtracted to obtain an estimate of the spillover effect, the additional effect on treated unit $i \in \mathcal{S}_{11}$ due to some proportion of $\mathcal{N}(i)$ being treated: $\hat{\tau}_{it}^{SE}= \hat{\tau}_{it}^{TE} - \hat{\tau}_{it}^{DE} = \hat{Y}_{it}^{\text{aug}}(0,1) - \hat{Y}_{it}^{\text{aug}}(0,0)$.
    % \begin{equation*}
    %     \label{eqn:2SE}
    %     \hat{\tau}_{it}^{SE}= \hat{\tau}_{it}^{TE} - \hat{\tau}_{it}^{DE} = \hat{Y}_{it}^{\text{aug}}(0,1) - \hat{Y}_{it}^{\text{aug}}(0,0)
    %     % &= \Bigg[\sum\limits_{j\in \{N_C+1,\ldots,N\}}{\gamma_{j}}^{(i)_{TE}} Y_{jt}\Bigg] - \Bigg[\sum\limits_{j \in (\{N_B+1, \ldots, N_C\} \setminus \mathcal{N}(i))}{\gamma_{j}}^{(i)_{DE}} Y_{jt} \Bigg]
    % \end{equation*}

\subsubsection{Inference}

Large-sample frequentist inferential methods are often unsuitable for inference in SCM due to the limited number of units, which render asymptotic approximations (based on $N\rightarrow\infty$) unreliable. Moreover, treatment is typically non-randomized in SCM applications, making randomization-based inference methods inapplicable. In addition, unlike traditional statistical settings where inference is often drawn for a population parameter, SCM aims to estimate the treatment effect for a single treated unit. As a result, inference in SCM has primarily relied on permutation-based methods, such as placebo tests, to assess significance \citep{abadieLH2021, abadie2010}. However, recent research has explored alternative inferential procedures for SCMs \citep{chernozhukov2024ttestsyntheticcontrols, li:2022, abadie2021, Cattaneo, chernozhukov2019, FirpoPossebom+2018}.

 % Traditional inference methods are not suitable for SCM due to small sample sizes and non-randomized treatment assignments. %There is no ``gold standard" inferential method for SCMs. 
 % Permutation-type arguments, such as placebo tests are commonly used in this setting \citep{abadie2010, abadieLH2021}. Ongoing research focuses on developing improved inference techniques for SCMs \citep{chernozhukov2019, FirpoPossebom+2018, li:2022, Cattaneo, chernozhukov2024ttestsyntheticcontrols}.

Herein, confidence intervals for ${\tau}_{it}^{TE}$ and ${\tau}_{it}^{DE}$ are constructed via conformal inference (\cite{ben-michael2021, chernozhukov2019}).The conformal inference procedure for Ridge ASCM consists of three key steps. First, for a given null hypothesis $H_0:\tau=\tau_0$, the observed post-treatment outcome $Y_{1T}$ at a single post-treatment time $t=T$ is adjusted by subtracting the hypothesized treatment effect $\tau_0$, yielding the adjusted outcome $\Tilde{Y}_{1T}=Y_{1T}-\tau_0$. %The dataset is then augmented to include this adjusted outcome for the treated unit. 
Second, Ridge ASCM is applied to the extended dataset, using $\Tilde{Y}_{1T}$ as the outcome for the treated unit at $t=T$. This yields synthetic control weights $\hat{\gamma}_i(\tau_0)$, which are then used to compute an estimate of the counterfactual post-treatment outcome of the treated unit, $\hat{Y}_{1T}(0)=\sum\hat{\gamma}_i(\tau_0)Y_{iT}$, and the time $T$ residual, $R_T=\Tilde{Y}_{1T}-\hat{Y}_{1T}(0)$. Finally, a p-value for $\tau_0$ is computed by comparing $R_T$ to the distribution of residuals obtained from pre-treatment periods: $p(\tau_0) = \frac{1}{T} \sum_{t=1}^{T_0} I \left\{ \left| Y_{1t} - \sum_i \hat{\gamma}_i(\tau_0) Y_{it} \right| \geq | R_T | \right\} + \frac{1}{T}$; i.e., the p-value is determined by the proportion of pre-treatment residuals that are larger than or equal to the post-treatment residual under the null hypothesis. The test is repeated over all possible null hypothesis values, and a $(1-\alpha)$-level confidence interval for $\tau$ is then constructed by $\text{CI}_{1-\alpha} = \{ \tau_0 : p(\tau_0) > \alpha \}.$ That is, the set of treatment effects $\tau_0$ that cannot be
rejected at a given significance level forms a confidence interval for the true treatment effect. By construction, these intervals have exact finite-sample coverage when the residuals are exchangeable across time periods. In cases where exchangeability is not guaranteed, \citet{chernozhukov2019} provide a finite-sample bound that ensures approximate validity as the number of pre-treatment periods increases.

\subsection{Estimation Error}

Ridge ASCM improves pre-treatment fit, which in turn reduces estimation error under different DGPs. In both autoregressive and linear factor models, estimation error can be decomposed into distinct components, with a common source being imbalance in lagged outcomes ($\mathbf{X}_1$ and $\mathbf{X}_0$, for now ignoring covariates). By allowing extrapolation via negative weights, Ridge ASCM improves pre-treatment fit relative to standard SCM, reducing bias at the cost of increased variance \citep{ben-michael2021}. This bias-variance trade-off is governed by the ridge penalty term $\lambda^\text{ridge}$, which adjusts the degree of extrapolation - higher values yield more conservative estimates, while lower values increase extrapolation but risk overfitting to noise. The selection of $\lambda^\text{ridge}$ is therefore critical in practice and can be optimized via cross-validation \citep{ben-michael2021}.

\subsubsection{Error Bounds under an Autoregressive Model}

Under an assumed autoregressive model, each unit's outcome at time \(T\) is modeled as a function of its own past outcomes. The error bound for ASCM-SC, the bound on the difference between the counterfactual post-treatment outcome and the Ridge ASCM estimator of the post-treatment outcome, is given by:
\[
\left| Y_{1T}(0) - \sum \hat{\gamma}_i^{\text{aug}} Y_{iT} \right| \leq \| \beta \|_2 \left\| \text{diag} \left( \frac{\lambda}{d_j^2 + \lambda} \right) (\Tilde{\mathbf{X}}_1 - \Tilde{\mathbf{X}}_0' \hat{\gamma}^{\text{scm}}) \right\|_2 + \delta \sigma (1 + \|\hat{\boldsymbol{\gamma}}^{\text{aug}}\|_2),
\]

\noindent where the first term represents imbalance in the lagged outcomes ($\mathbf{X}_1$ and $\mathbf{X}_0$), when the synthetic control fails to fully match the treated unit's pre-treatment outcomes, and the second term represents residual noise, which accounts for any random noise in the post-treatment period that cannot be explained by past outcomes. The overall error decreases with better pre-treatment fit, with more lagged outcomes improving the model's accuracy. The error is bounded by the imbalance between ($\mathbf{X}_1$ and $\mathbf{X}_0$) and the $L^2$ norm of the Ridge ASCM weights \citep{ben-michael2021}.

\subsubsection{Error Bounds under a Linear Factor Model}

When outcomes follow a linear factor model, they are influenced by unobserved latent factors that vary across units and over time. This introduces additional sources of estimation error beyond those in the autoregressive model. Specifically, the error bound includes terms accounting for: (1) imbalance in the lagged outcomes, (2) excess approximation error, which can also be described as a measure of extrapolation when implementing ASCM vs. SCM, (3) SCM approximation error, and (4) post-treatment noise. A formal definition of this bound can be found in Theorem 1 of \citet{ben-michael2021}. In the case where perfect pre-treatment fit is achieved, only the latter 2 components are considered, as in \citet{abadie2010}. The addition of the first two components provides the worst-case error bound in which a control unit given a large weight is only similar to the treated unit due to noise. When covariates are incorporated, an additional component of error can arise due to imbalance in the covariates, but this can be bounded and can be eliminated under certain assumed relationships between the outcome and the covariates \citep{ben-michael2021}. In both the linear factor and autoregressive models, error bounds improve as more pre-treatment periods are available and as the synthetic control provides better pre-treatment balance

\subsubsection{Impact of Interference on Estimation Error}
The error bounds described above assume that errors are independent across units and time. Thus, in settings with interference, this assumption may be violated. %if the outcome of one unit is influenced by the treatment status of another. This can introduce additional bias by breaking the independence structure required for valid bounds \citep{ben-michael2021}. 
However, by implementing ASCM-SC under the assumption that the potential outcome of unit $i$ with neighbors $\mathcal{N}(i)$ is given by $Y_{it}(a,q)$ and leveraging stratified control groups - units in $\mathcal{S}_{01}$ and $\mathcal{S}_{00}$ that are assumed to be unaffected by the treated unit - we preserve the independence assumption across units. This ensures that the bias bounds described above for the autoregressive and linear factor models remain valid for the ASCM-SC direct and total effect estimators.

\section{Simulations} \label{sec:chp3_sims}

\subsection{ASCM with stratified controls vs. all controls}\label{sec:chp3_sim_1}

Extensive simulation studies were conducted, calibrated to our empirical illustration in Section \ref{sec:chp3_applied_chile}, to assess the performance of ASCM-SC under various DGPs: a linear factor model, a simplified linear factor model with only unit- and time-fixed effects, and an autoregressive model. Results of the substantial gains from ASCM as compared to SCM in null simulations under these DGPs in the absence of interference can be found in \citet{ben-michael2021}. Herein, simulations include terms for the direct effect of treatment as well as the spillover effect due to some portion of neighbors being treated. For the linear factor models and simplified linear factor models, three different DGPs were simulated for each: (i) with no covariates, (ii) with limited covariates, including those used to determine treatment, and (iii) with all covariates in the dataset. See Web Appendix A for further simulation details.

For each DGP, the bias and coverage properties of the ASCM-SC estimators of direct, total, and indirect effects were examined, and the bias and coverage properties of the ASCM-SC direct effect estimator and Ridge ASCM using all available control units (hereafter referred to as `naive ASCM') were compared. Each of the four estimators (direct, total, indirect, and naive) was implemented 3 times: including no covariates, limited covariates, and all covariates in the estimation process. 

\begin{figure}[h]
\centering
\includegraphics[width=1\linewidth]{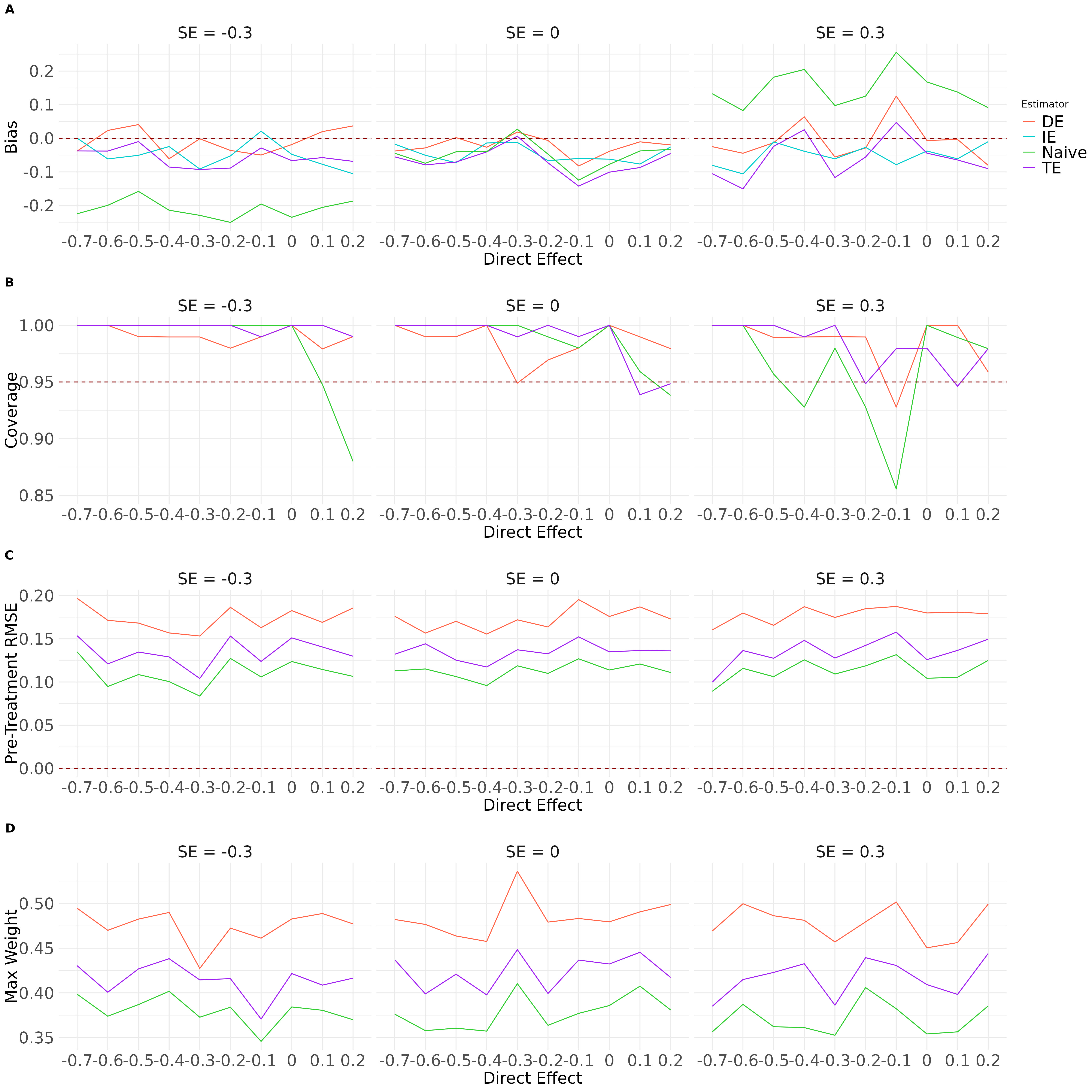}
\caption{Simulation results for a linear factor model with limited covariates, estimating direct, total, and indirect effects of treatment over varying direct effects and indirect effects (listed above each plot) when utilizing limited covariates in the ASCM estimation process. Column (A) Bias results for all 4 estimators, (B) Coverage of estimators, (C) Pre-treatment root mean square error of estimators, (D) Maximum weight assigned to a control unit for all estimators.
\newline DE = Direct Effect, IE = Indirect Effect, TE = Total Effect} 
\label{fig:Paper2_Fig1}
%\caption*{}

\end{figure}

Below, results of the linear factor model with limited covariates included in both the DGP and the estimation process are presented (Figure \ref{fig:Paper2_Fig1}).  Results from other DGPs exhibit similar bias and coverage patterns and can be found in Web Appendix A. %of the \nameref{Supp}. 
The direct effect in the DGP was varied from -0.7 to 0.2 and the indirect effect from -0.3 to 0.3; with the outcome being $log(R_t)$, this varies between an approximate 50\% decrease up to an approximate 35\% increase in $R_t$ due to lockdown, and an approximate 25\% decrease up to an approximate 35\% increase in $R_t$ due to a neighbor being under lockdown. Figure \ref{fig:Paper2_Fig1} shows the bias, coverage, pre-treatment root mean squared error (RMSE), and maximum weight assigned to a control unit for each estimator for a selection of these scenarios. All scenarios (which exhibit similar patterns) are in Web Appendix A. %of the \nameref{Supp}. 

% and separately for the direct, total (and hence indirect) effects, we compare the results of four estimators: (i) ASCM-SC, (ii) ASCM-SC including limited auxiliary covariates in parallel to lagged outcomes, (iii) ASCM-SC including all auxiliary covariates in parallel to lagged outcomes, and (iv) Ridge ASCM using all available control units (hereafter referred to as `naive ASCM'). See Appendix for further simulation details and extensive results.

These simulations highlight a few key results. First, in general, the bias of the direct effect estimate is smaller when using ASCM-SC compared to naive ASCM except for in simulations with no spillover effect, and the bias of naive ASCM tends to increase in scenarios with larger indirect effects (Figure \ref{fig:Paper2_Fig1}, DE \& Naive). Second, across all estimators, utilizing conformal inference prediction intervals typically results in at least 95\% coverage of the true effect. Third, overall, the pre-treatment fit of ASCM-SC is not as good as naive ASCM but is still small relative to the outcome values. Finally, the maximum weight assigned to an individual control unit is, on average, larger when using ASCM-SC as compared to naive ASCM, but synthetic controls under ASCM-SC are still a weighted combination of various control units (i.e., there is not strong evidence of overfitting). However, the distribution of weights to control units in settings with a relatively small number of controls should be monitored to ensure overfitting does not occur (i.e., the synthetic control is simply a weighted combination of a very small number of control units).

\subsection{ASCM with stratified controls vs. iSCM}

In addition to comparing ASCM-SC to naive ASCM, we also assessed its bias properties in estimating the direct effect relative to \citet{distefano2024}'s iSCM method, which can account for multiple treated units when interference is plausible. ASCM-SC demonstrated lower bias in estimating direct effects compared to iSCM. Results for the linear factor model with no covariates are shown in Figure 2. Across varying direct and indirect effects, ASCM-SC remains consistently close to zero bias, while iSCM exhibits increased bias as the spillover effect moves further from zero. This suggests that ASCM-SC provides more consistent and reliable direct effect estimates in settings with multiple treated units that may have spillover effects, reinforcing its utility for policy evaluations in such contexts.

% In addition to comparing ASCM-SC to naive ASCM, the bias properties of ASCM-SC were also compared to two SCM-based interference-related methods applicable in a setting with multiple treated units. Particularly, we compare the bias properties of ASCM-SC in estimating the direct effect to \citet{distefano2024}'s iSCM method and \citet{cao_dowd2019}'s method.  

% ASCM-SC demonstrated lower bias in estimating direct effects compared to iSCM. Results for the linear factor model with no covariates are shown in Figure \ref{paper2_Fig2}. Across varying direct and indirect effects, ASCM-SC remains consistently close to zero bias, while iSCM exhibits increased bias as the spillover effect moves further from zero. This suggests that ASCM-SC provides more consistent and reliable direct effect estimates in this setting of several clustered treated units that may have spillover effects on one another, reinforcing its utility for policy evaluations involving multiple treated units.

\begin{figure}[h]
\centering
\includegraphics[width=0.9\linewidth]{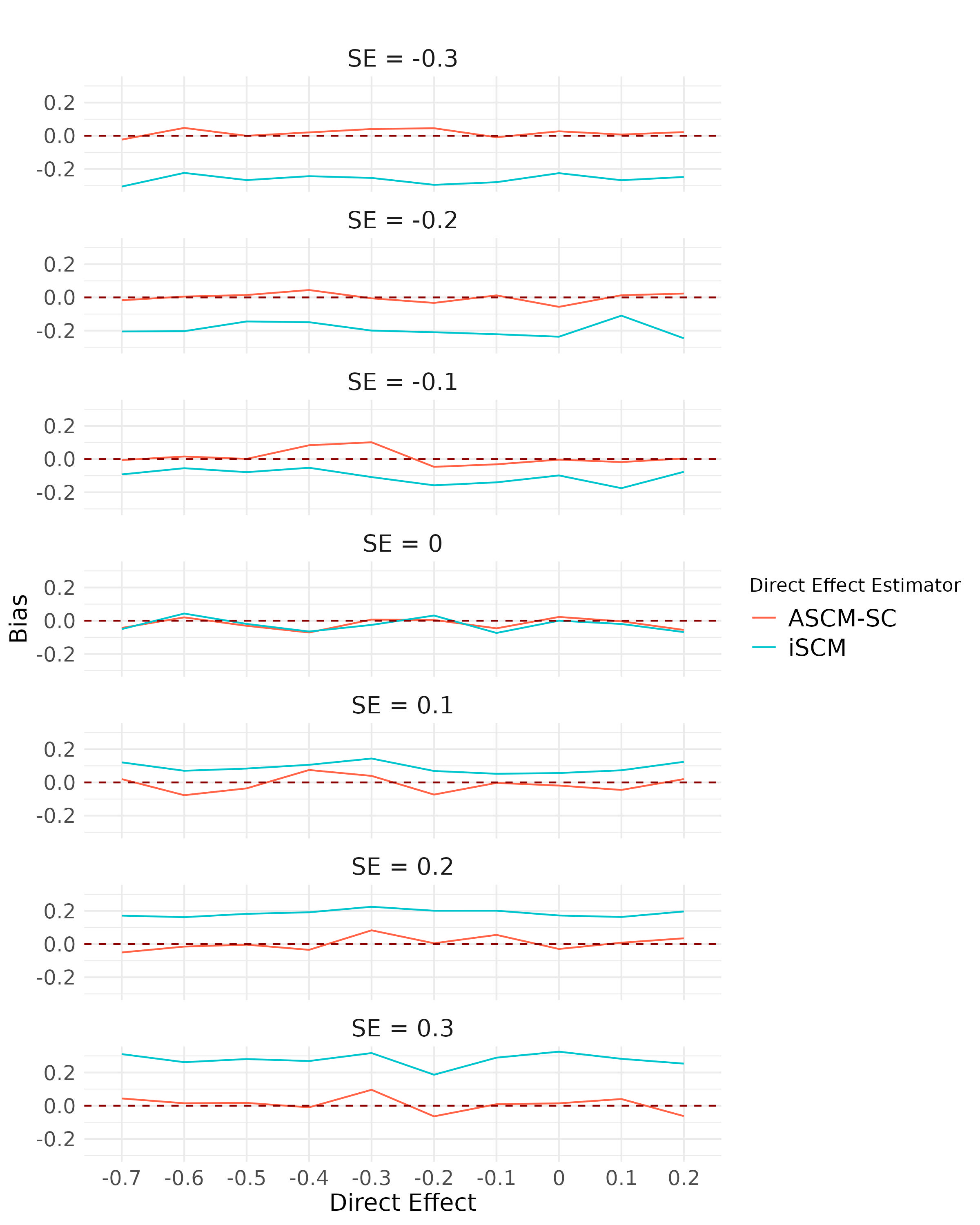}
\caption{Simulation results for a simplified linear factor model with no covariates, estimating direct effects of treatment over varying direct effects and indirect effects (listed above each plot) compared to the direct effect estimated via \citet{distefano2024}'s iSCM method. Bias results are shown for both estimators.
\newline DE = Direct Effect}
\label{fig:paper2_Fig2}
\end{figure}

\subsection{ASCM with stratified controls vs. \citet{li:2022}'s method} \label{sec:chp3_Yige}
\begin{figure}[h]
\centering
\includegraphics[width=0.8\linewidth]{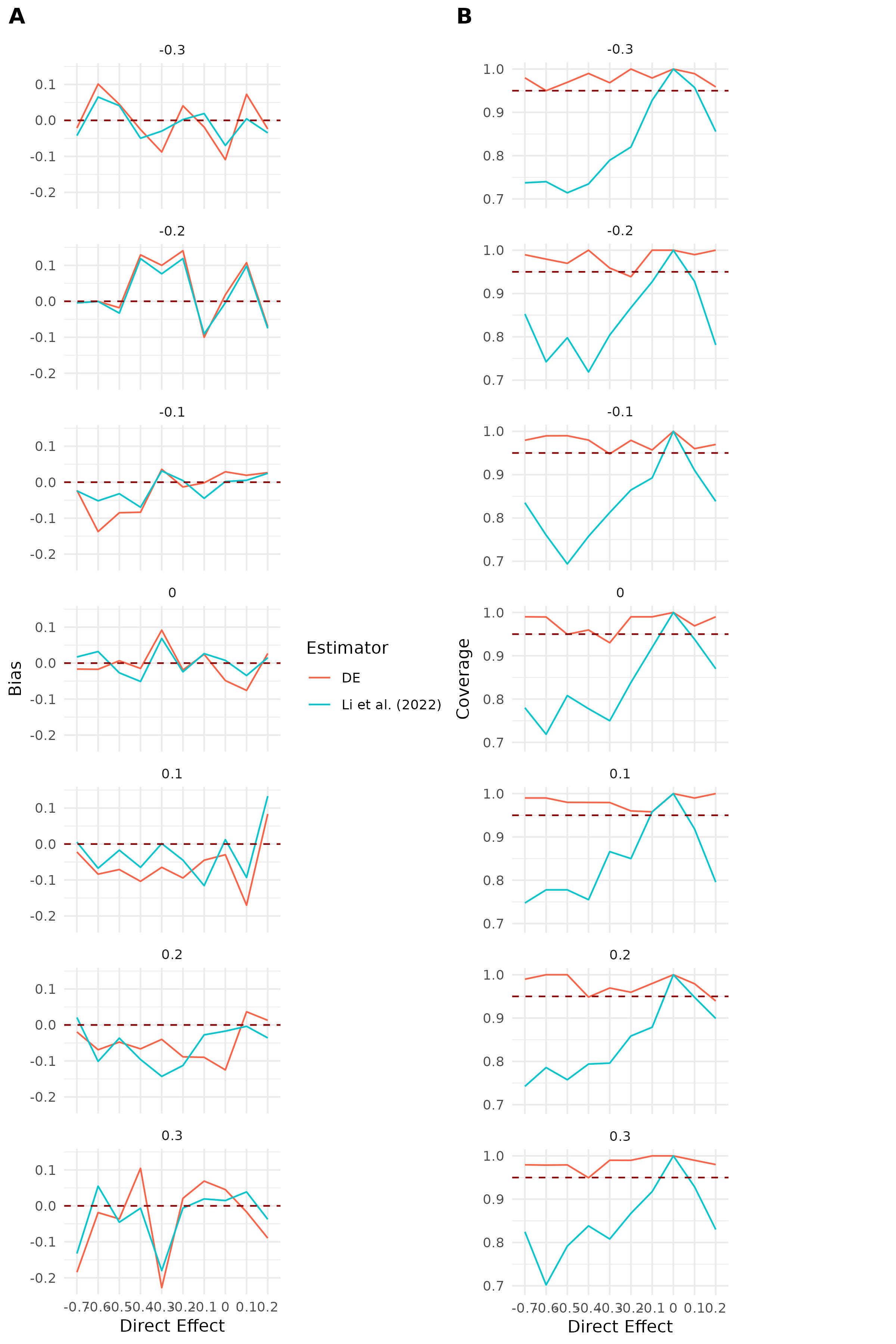} 
\caption{Simulation results for a simplified linear factor model with all covariates, estimating direct effects of treatment over varying direct effects and indirect effects (listed above each plot) when utilizing all covariates in the ASCM estimation process, compared to the direct effect estimated via \citet{li:2022}'s method. Column (A) Bias results for both estimators, (B) Coverage of estimators.}\label{fig:Yige_main}
\end{figure}

%In addition to the above-mentioned methods that incorporate interference into synthetic control methodology, 
ASCM-SC was also compared to the approach in \citet{li:2022}. Utilizing the same dataset described in Section \ref{sec:chp3_data}, \citet{li:2022} estimated the impact of localized lockdowns on the transmission of COVID-19 in Chile implementing a modification of ASCM that allowed for spillover by modifying the dataset to extend lockdowns and creating copies of control comunas. Specifically, they investigated the effect of extending lockdown periods within the municipality of interest as well as in neighboring municipalities. Their pre-intervention period consisted of the last week of lockdown during the first wave of lockdowns, and they aimed to estimate the effect of extending the lockdown rather than lifting it. More details of their approach and comparison to ASCM-SC can be found in Web Appendix A. Figure \ref{fig:Yige_main} shows the results of \citet{li:2022}'s method as compared to ASCM-SC on the simulated data (as described in Section \ref{sec:chp3_sim_1})  when applied to estimate the effect of going under lockdown under a simplified linear factor model with all covariates. We found that their method resulted in potential overfitting in the pre-treatment period and undercoverage in the post-treatment period; results from other DGPs showed similar patterns.

\section{Effects of Comuna Level COVID-19 Lockdowns in Chile}\label{sec:chp3_applied_chile}

In this section, the data from \citet{li:2022} is utilized to estimate the effects of comuna-level lockdowns, specifically focusing on the initial wave of lockdowns that began at the end of March 2020. While \citet{li:2022} analyzed the effects of varying lockdown durations and the influence of neighboring municipalities, focusing on lockdown extension as the intervention of interest, this analysis treats the initial lockdown implementation as the primary intervention of interest. Accordingly, the pre-intervention period consists of the week prior to the first wave of lockdowns, and ASCM-SC is implemented to estimate the direct effect of a comuna going under lockdown as well as the total effect when additionally accounting for the effect of at least one neighboring comuna going under lockdown. By implementing ASCM-SC, we aim to provide a more nuanced understanding of how these early lockdowns influenced the trajectory of COVID-19 transmission across different comunas in Chile.

 Let $Y_{it}$ be the instantaneous reproduction number in municipality $i$ at time $t$. Let $\mathbf{X}_i$ contain the seven-day history of: lockdown interventions in municipality $i$, the proportion of the population in neighboring municipalities under lockdown, and the instantaneous reproduction number in municipality $i$. Additionally, $\mathbf{X}_i$ contains several municipality-level characteristics that may affect virus transmission: the proportion of the population who were female, living in rural areas, older than 65, living below the official poverty level, living in overcrowded households, and lacking adequate sanitation infrastructure, as well as average monthly income.

% ASCM-SC was applied to assess the impact of localized lockdowns in the seven comunas in Chile that went under lockdown during the initial lockdown period, including Independencia, Las Condes, Lo Barnechea, Nunoa, Providencia, Santiago, and Vitacura. For each comuna, ASCM-SC was implemented to estimate, separately, the direct and total effects of lockdown, as described in Section 2, and 95\% point-wise confidence intervals were computed via conformal inference. Figures 2 through 4, first row,hc show the $log(R_t)$ values for Lo Barnechea, Providencia, and Santiago and their synthetic controls (direct effect synthetic control in column B, total effect in column C). We chose these three comunas for reasons indicated in \citet{li:2022}: they all have a lot of interdependence with other municipalities in Greater Santiago but represent different types of comunas. Santiago is the nation’s center for financial, commercial, and political activities, while Providencia, a more affluent urban municipality, is known for its commercial significance and a notably older population. Lo Barnechea, by contrast, is largely residential with a diverse population and minimal commercial development \citep{li:2022}. Results for the remaining four comunas along with balance tables for pre-treatment covariates and lagged outcome variables for all seven comunas can be found in Web Appendix B of the \nameref{Supp}.

\begin{figure}[H]
\centering
\includegraphics[width=1\linewidth]{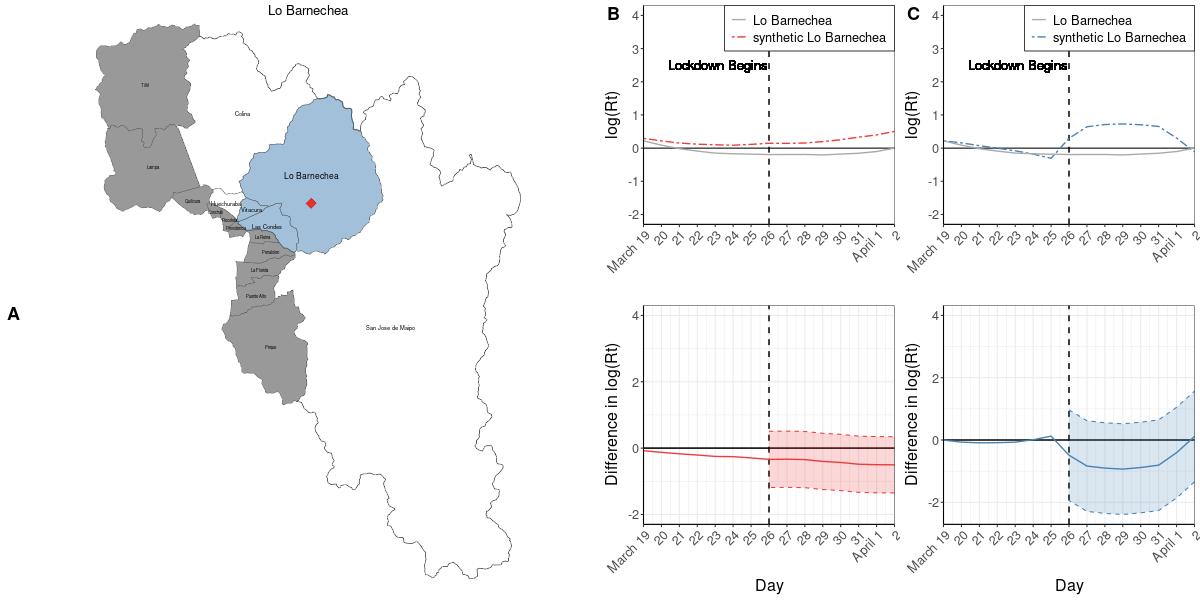}
\caption{Estimated effects of lockdown in Lo Barnechea utilizing ASCM-SC. $log(R_t)$, log of the average number of secondary COVID-19 cases per primary infected case, for Lo Barnechea and synthetic Lo Barnechea are shown in the top row for (B) the direct effect and (C) the total effect. Gap plots of Lo Barnechea versus synthetic Lo Barnechea are shown in the bottom row. Vertical line indicates the start of the first lockdown wave.} \label{fig:LB}
\end{figure}

\begin{figure}[H]
\centering
\includegraphics[width=1\linewidth]{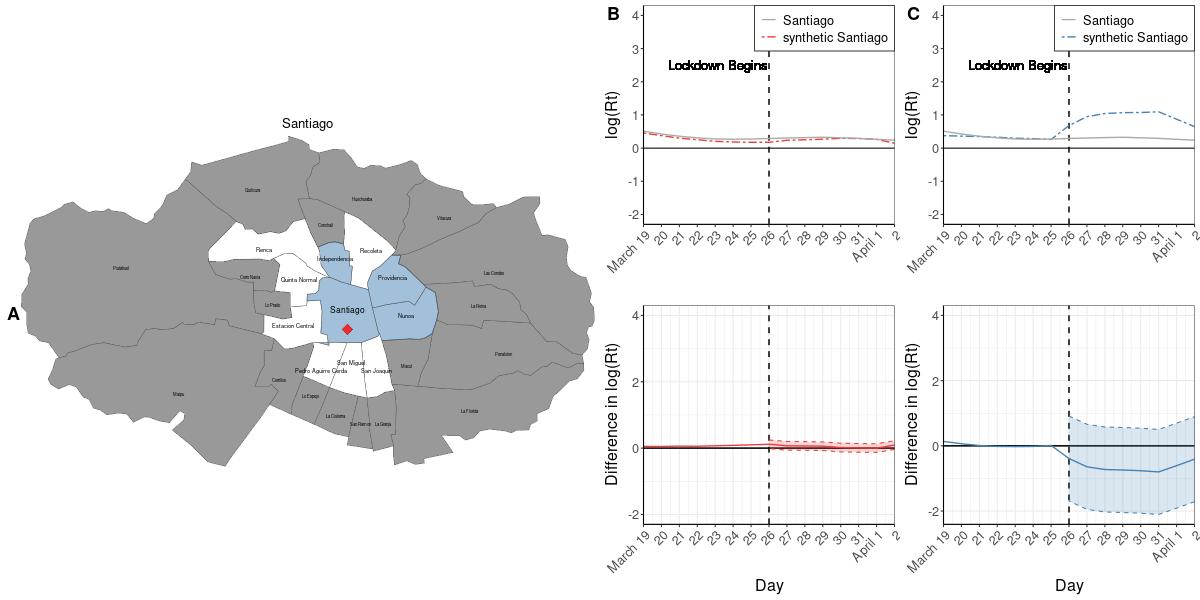}
\caption{Estimated effects of lockdown in Santiago utilizing ASCM-SC.  $log(R_t)$, log of the average number of secondary COVID-19 cases per primary infected case, for Santiago and synthetic Santiago are shown in the top row for (B) the direct effect and (C) the total effect. Gap plots of Santiago versus synthetic Santiago are shown in the bottom row. Vertical line indicates the start of the first lockdown wave.}\label{fig:Santiago}
\end{figure}

ASCM-SC was applied to assess the impact of localized lockdowns in the seven comunas in Chile that went under lockdown during the initial lockdown period, including Independencia, Las Condes, Lo Barnechea, \~Nu\~noa, Providencia, Santiago, and Vitacura. For each comuna, ASCM-SC was implemented to estimate, separately, the direct and total effects of lockdown, as described in Section \ref{sec:paper2_methods}, and 95\% point-wise confidence intervals were computed via conformal inference. Figures \ref{fig:LB} and \ref{fig:Santiago}, first row, show the $log(R_t)$ values for Lo Barnechea and Santiago and their synthetic controls (direct effect synthetic control in column B, total effect in column C). These two comunas are highlighted for reasons indicated in \citet{li:2022}: they both have a lot of interdependence with other municipalities in Greater Santiago but represent different types of comunas. Santiago is the nation's center for financial, commercial, and political activities. Lo Barnechea, by contrast, is largely residential with a diverse population and minimal commercial development. Results for the remaining five comunas along with balance tables for pre-treatment covariates and lagged outcome variables for all seven comunas can be found in Web Appendix B. %Web Appendix B of the \nameref{Supp}.

In Lo Barnechea, there is an estimated direct effect equating to an approximate 30 - 40\% reduction in the average number of secondary cases per primary case, with a greater difference between Lo Barnechea and its synthetic control over time (Figure \ref{fig:LB}). Accounting for the fact that at least one of Lo Barnechea's neighbors also went under lockdown reveals a greater reduction in the average number of secondary cases per primary case, equating to an approximate 33 - 60\% reduction, although this difference wanes over the course of the week of lockdown. This indicates an estimated additional reduction in $R_t$ of between approximately 15\% and 40\% over the first five days of lockdown due to a neighbor being under lockdown, with the effect waning over time. However, although the estimated effects suggest a potential reduction in transmission, at least during the initial five days of lockdown, all confidence intervals for these estimates include zero, indicating a lack of strong evidence that going under lockdown nor having a neighbor under lockdown significantly impacted transmission rates in Lo Barnechea.

% In Providencia, we see an early reduction in transmission of about 20\% in both the direct and total effect estimation, indicating little influence of a neighbor being under lockdown (Figure 3). This effect weakened over time, with the gap between observed and synthetic $log(R_t)$ gradually closing in estimating the direct effect and diverging when estimating the total effect, indicating a direct effect equating to a potential increase in transmission for the latter part of the first week of lockdown, and an even larger increase in transmission when accounting for neighbors being under lockdown. As in Providencia, all confidence intervals for these estimates include zero, and thus we don't have strong evidence that going under lockdown nor having a neighbor under lockdown significantly impacted transmission rates in Providencia.

Results in Santiago show a nearly null direct effect of lockdown throughout the first week (Figure \ref{fig:Santiago}). However, the estimated total effect indicates a reduction in transmission throughout the first week of lockdown, indicating a reduction in transmission as a result of neighbor(s) also being under lockdown. This total reduction in transmission ranged from 32\% to 55\% over the first week of lockdown. However, as in the previous comunas, all confidence intervals include zero, and thus it is possible that the lockdowns, either directly or indirectly, did not have a strong impact on transmission.

% \begin{figure}[H]
% \centering
% \includegraphics[width=1\linewidth]{figures/map_plot_Providencia.jpeg}
% \caption{Results of lockdown lift in Providencia with the same analysis in Figure 2. $log(R_t)$, log of the average number of secondary COVID-19 cases per primary infected case, for Providencia and synthetic Providencia are shown in the top row for (B) the direct effect and (C) the total effect. Gap plots of Providencia versus synthetic Providencia are shown in the bottom row. Vertical line indicates the start of the first lockdown wave.}
% \end{figure}

Across the seven comunas that went under lockdown in the initial lockdown period beginning late March, 2020, in some cases, the lockdowns were associated with a noticeable decrease in transmission, evidenced by a lower $log(R_t)$ in the treated comuna compared to its synthetic counterpart. However, in other comunas, the decrease in transmission was less pronounced, or the effect was mixed, with some initial reduction followed by a rebound or stabilization of transmission rates. The majority of confidence intervals for both direct and total effects included zero, suggesting a lack of strong evidence of lockdown effectiveness, either directly or indirectly, in impacting the transmission of COVID-19 as measured by $R_t$.

\section{Conclusion}

%\subsection{Summary}
Synthetic control methods have become an increasingly popular approach for estimating the effect of an intervention on a single treated unit. However, the majority of synthetic control-based methods assume no spillover of treatment effect from one unit to another. Ignoring potential spillover effects can lead to biased estimates of the treatment effect. To address this issue, ASCM was modified by stratifying the control group to separately estimate the direct and total, and in turn indirect, effects of an intervention in settings where several spatially clustered units receive treatment.
%when implemented in a setting in which several units receive treatment, and the units receiving treatment are spatially clustered. 
This method results in smaller bias of the direct effect estimate and allows for teasing apart the direct and indirect effects of an intervention. 

%\subsection{Limitations}
When implementing ASCM-SC, one must take note of the number of available controls in each stratified group. If one of the stratified control groups has too few controls, this could lead to overfitting of the synthetic control to a single or few treated units. Additionally, regardless of the number of controls in the stratified control groups, careful attention should be paid to the weight distribution assigned to the controls to ensure no overfitting.

%\subsection{Future Directions}
 There is no gold standard for statistical inference procedures for synthetic control estimates, and they remain an active area of research. As noted in Section \ref{sec:chp3_sims}, utilizing conformal inference for confidence interval estimation when implementing ASCM-SC often leads to overcoverage. Future work could focus on increasing the precision of confidence intervals for synthetic control estimates, both with and without interference present.

\backmatter

%%%%%% include this section if you wish to acknowledge people,
%%%%%% grant support, etc.

\section*{Acknowledgements}

 The authors would like to thank Jos\'e Zubizarreta and Yige Li for their work creating the Chile COVID-19 dataset and for their helpful contributions to the analysis of comuna level COVID-19 lockdowns in Chile. This work was supported in part by NIH grant R01 AI085073 and by NIEHS grant T32 ES007018.
 \vspace*{-8pt}

 \section*{Data Availability Statement}
 The data that support the findings of this study are openly available at \newline https://scholar.harvard.edu/zubizarreta/code-and-replication-files.

% The Duchenne Natural History Study (DNHS) data that support the findings in this paper are available on request. Please email info@trinds.com for more details. The data are not publicly available due to privacy or ethical restrictions.

%%%%%% include this section only if your manuscript refers to supplementary
%%%%%% materials -- see Instructions for Authors at 
%%%%%% http://www.tibs.org/biometrics

\section*{Supplementary Materials}\label{Supp}

Web Appendix A referenced in Section~\ref{sec:chp3_sims}  and Web Appendix B referenced in Section~\ref{sec:chp3_applied_chile} are available upon request.
%with this paper at the Biometrics website on Wiley Online Library.
\vspace*{-8pt}

\label{lastpage}

\end{document}